\documentclass[11pt,english]{article}
\usepackage[T1]{fontenc}
\usepackage[latin9]{inputenc}
\usepackage{amsmath}
\usepackage{amssymb}

\makeatletter
\newcommand{\be}{\begin{equation}}\newcommand{\ee}{\end{equation}}\newcommand{\bea}{\begin{eqnarray}}\newcommand{\eea}{\end{eqnarray}}

\def\({\left(}\def\){\right)}\usepackage{epsfig}

\makeatother

\usepackage{babel}

\begin{document}

\title{Flux-conserving hyperbolic systems and two-dimensional evaporating black holes}

\author{Amos Ori \\ Department of Physics, Technion-Israel Institute of Technology, \\  Haifa 32000,  Israel }

 \maketitle
 
 \begin{abstract} 
 A class of semi-linear hyperbolic systems 
 in 1+1 dimensions was investigated several years ago by Ori and Gorbonos \cite{gorbonos}. This class, to which we shall refer as {\it flux-conserving systems}, exhibits a variety of interesting mathematical properties. 
Here we demonstrate how the formalism of flux-conserving systems can be applied to the problem of black-hole evaporation in 1+1-dimensions. More specifically, we show how the semiclassical CGHS \cite{CGHS} field equations may be approximated by a certain flux-conserving system. 
\end{abstract}


\section{Introduction}
The investigation of gravitational collapse and black-hole (BH) dynamics often leads to systems of partial differential equations (PDEs) in 1+1 dimensions.
This happens, for example, in the study of four-dimensional (4D) BHs when spherical symmetry is assumed. The usage of spherical symmetry for dimensional reduction in BH physics has been a traditional research strategy in General Relativity for many decades---since the discovery of the Schwarzschild and Reissner-Nordstrom solutions almost a century ago. 

About two decades ago Callan, Giddings, Harvey and Strominger (CGHS) \cite{CGHS} introduced a two-dimensional toy model for investigating the semiclassical evaporation of black holes, with the hope of addressing the information puzzle. 
For appropriate initial conditions, this model describes the formation of a BH as well as its subsequent evaporation. Mathematically, this model again reduces to a system of PDEs in 1+1 dimensions. 

Among the family of 4D spherically-symmetric solutions, an interesting subclass is the one in which the matter content is made of (at most) two radial fluxes of null fluids. The null fluxes are assumed to be pressure-less and non-interacting (except gravitationally), hence each flux is conserved separately. If only one such null fluid is present, 
the system admits an exact solution, known as the Vaidya solution \cite{vaidya}. Obviously, if both fluxes vanish the system reduces to the Schwarzschild solution.  It is useful to extend this subclass by allowing the presence of a spherically-symmetric electric field, in addition to the null fluids. This gives rise to the mass-inflation model \cite{poisson}, 
the charged Vaidya solution \cite{vaidya2} (in the case where one flux vanishes), and, obviously, to the Reissner-Nordstrom solution (in the special case where both fluxes vanish). All these solutions played a central role in studying the non-trivial internal structure of spherically-symmetric charged BHs \cite{graves}, and particularly in investigating the singularity which develops at the inner horizon \cite{Hiscock,poisson,Ori1}. 

A few years ago Ori and Gorbonos \cite{gorbonos} explored a simple class of semi-linear hyperbolic systems in 1+1 dimensions, to which I will refer as {\it flux-conserving systems}. Each such system is characterized by a single function of one variable, denoted by $h(\bar{R})$  below. This class, 
which was first introduced in Ref. \cite{Ori-charged}, 
exhibits several interesting mathematical properties. In particular, in any such system one can identify two conserved null fluxes. In the special case where one of these fluxes vanish, the solution takes a Vaidya-like form and the system then reduces to an ordinary differential equation (ODE), which may remarkably simplify the analysis. The flux-conserving systems contain, as special cases, the conventional Vaidya \cite{vaidya}
and mass-inflation \cite{poisson} solutions (as well as the Schwarzschild and Reissner-Nordstrom solutions), which all correspond to the same function $h(\bar{R})$ \cite{gorbonos}. It also includes the {\it classical} CGHS solution, 
though the semiclassically-corrected CGHS system does not belong to this class. 

The original motivation for introducing these semi-linear systems in Ref. \cite{gorbonos} was to provide a simpler (and at the same time wider) mathematical context for investigating the formation of null weak singularities at the inner horizon of BHs. This goal was undertaken 
by Gorbonos and Wolansky  \cite{second}, who
demonstrated that for a certain (piecewise-linear) class of functions $h(\bar{R})$ a generic null weak singularity indeed forms. 

The present paper will focus on a different aspect of these semi-linear systems: the very conservation of the individual fluxes,
and its possible application to analyzing the evolution of dynamical BHs. 
As an example, consider a flux-conserving system which describes a BH bombarded by ingoing radiation (with no outflux). In such a system 
the influx at past null infinity directly determines the rate of change of the BH mass. In the traditional framework of conventional 4D spherically-symmetric BHs this is illustrated by the ingoing Vaidya solution, in which the information is fully encoded in the mass-function $M(v)$. The same applies to (Vaidya-like solutions for) evolving BHs in {\it any} type of flux-conserving system. 

More specifically, 
our main goal in this paper is to demonstrate how the mathematical concept of flux-conserving systems may be used to analyze the evaporation of two-dimensional semiclassical BHs. 

As was already mentioned above, the classical CGHS system is a flux-conserving one \cite{gorbonos}, but this situation changes when the semiclassical effects are included. 
At first glance, an evaporating BH (say, in the Hartle-Hawking state) may actually be regarded as an archetype of a system which violates flux-conservation: No influx arrives from past null infinity, and yet negative influx penetrates into the BH and causes its evaporation. We shall show here, however, that by a re-definition of the field variables, the CGHS field equations may be transformed into a form which is approximately flux-conserving. 
More specifically, as long as the evaporating BH is still macroscopic, one may expand the semiclassical field equations in a small quantity (proportional to the number $N$ of quantum fields and inversely proportional to the variable $R$ defined below). The zero order in this expansion yields the classical system, whereas the next order represents the leading semiclassical effects. We shall show that up to (and including) this first order, in our new variables the transformed field equations constitute a flux-conserving system. This provides a powerful analytic tool for analyzing the structure of the semiclassical spacetime of a CGHS evaporating BH. 

We start with a very brief review of the formalism of flux-conserving systems (Sec. II) and the CGHS model (Sec. III)---introducing our own notation for the latter, and focusing on the properties which are relevant to the present analysis. Then in Sec IV we perform the field redefinitions, which eliminate the first-order derivative terms from the evolution equations. Finally in Sec. V we explore the large-$R$ approximate system (which incorporates the leading-order semiclassical effects), and demonstrate that it constitutes a flux-conserving system. We then describe how the CGHS evaporating-BH spacetime may be obtained (approximately) from an ingoing Vaidya-like solution of this system.

\section{Flux conserving systems}

By {\it flux-conserving system} we refer to a semi-linear hyperbolic system
in $1+1$ dimensions, for two unknowns $\bar{R}(u,v)$ and $\bar{S}(u,v)$,
which takes the form 
\begin{equation}
\bar{R}_{,uv}=e^{\bar{S}}F(\bar{R})\ ,\ \ \ \bar{S}_{,uv}=e^{\bar{S}}F'(\bar{R}),
\label{evf}
\end{equation}
for any given function $F(\bar{R})$, where a prime denotes $d/d\bar{R}$. 

This class of hyperbolic systems displays several unique mathematical
properties, which were discussed in some detail in Ref. \cite{gorbonos}. 
\footnote{The field variables are denoted $R,S$ therein, instead of $\bar{R},\bar{S}.$
Here we use the over-bar variables to distinguish them from the specific variables
of the CGHS model presented below. Similarly, we use $\bar{T}$
below to distinguish it from the physical stress-energy tensor $T$, and
the same for the mass parameter $\bar{m}$ and the mass function $\bar{M}(v)$.} 
Here we shall briefly mention some of these properties. 

First, the system admits a gauge symmetry: If one transforms the coordinates 
as $u\rightarrow\tilde{u}(u),v\rightarrow\tilde{v}(v)$,
then $\bar{R}$ behaves as a scalar (it is unchanged), but $\bar{S}$
transforms as 
\begin{equation}
\tilde{S}=\bar{S}-\ln\left(\frac{d\tilde{v}}{dv}\right)-\ln\left(\frac{d\tilde{u}}{du}\right)
\label{coo}
\end{equation}
(that is, $e^{\bar{S}}$ transforms like a metric component $g_{uv}.$) 

Second, we define the two flux functions \begin{equation}
\bar{T}_{uu}\equiv\bar{R}_{,u}\bar{S}_{,u}-\bar{R}_{,uu}\:,\;\;\bar{T}_{vv}\equiv\bar{R}_{,v}\bar{S}_{,v}-\bar{R}_{,vv}\label{eq:fluxes}\end{equation}
where a comma denotes partial dervative. 
It then follows from the hyperbolic system (\ref{evf}) that 
\begin{equation}
\bar{T}_{uu,v}=\bar{T}_{vv,u}=0.
\end{equation}
Namely, $\bar{T}_{uu}$ only depends on $u$, and $\bar{T}_{vv}$
anly depends of $v$. We refer to this property as {\it conservation of
fluxes}. 

Naturally the solutions of this system may be divided into three categories:
(i) solutions 
with $\bar{T}_{uu}=\bar{T}_{vv}=0$,
(ii) solutions in which only one flux vanishes; 
and (iii) solutions in which both
fluxes are non-vanishing. We refer to case (i) as {\it vacuum-like},
and to case (ii) as {\it Vidya-like}. Case (iii) is the mathematical generalization
of the 4D spherically-symmetric systems with two null fluids, for example the one studied in Ref. \cite{poisson}. 

Next we define the "generating function'' $h(\bar{R})$ by 
\begin{equation}
F(\bar{R})=-h'(\bar{R}). 
\label{hFromF}
\end{equation}
More precisely this is a one-parameter family of functions, due to
the arbitrary integration constant. It will often be convenient to express
this free parameter explicitly in the form 
\begin{equation}
h(\bar{R})=h_{0}(\bar{R})-\bar{m}, 
\label{hint}
\end{equation}
where $h_{0}(\bar{R})$ denotes a {\it specific} integral of (\ref{hFromF}), and $\bar{m}$ is an arbitrary constant. 

In the vacuum-like case (i) the space of solutions reduces to a one-parameter
family of solutions (apart from gauge transformations), characterized
by their "mass parameter" $\bar{m}$. It is convenient to express
the vacuum-like solutions in an "Eddington-like'' gauge, in which both $\bar{R}$
and $\bar{S}$ depend on $x\equiv v-u$ solely. For given $\bar{m}$,
$\bar{R}(x)$ is then determined from the ODE 
\begin{equation} 
\frac{d\bar{R}}{dx}=h_{0}(\bar{R})-\bar{m},
\end{equation} 
and then $\bar{S}(x)$ by 
$e^{\bar{S}}=h_{0}(\bar{R})-\bar{m}$. 
The Eddington-like gauge thus exhibits the static nature of the vacuum-like solutions. 
\footnote{In cases where the vacuum-like solution corresponds to a {}``black hole'' (see below), this staticity characterizes the external region. Inside the horizon, the vacuum-like solution is homogeneous rather than static. The functions $\bar{R},\bar{S}$ then depend on $t\equiv v+u$. } 

Assume now that $h(\bar{R})$ is monotonically increasing in some range of $\bar{R}$.  
\footnote{This may always be arranged, as long as $h(\bar{R})$ is not constant, by transforming $\bar{R}\rightarrow-\bar{R}$ if necessary (while fixing $h$).} 
Then, at least for some range of the parameter $\bar{m}$, $h$ will vanish at a certain $\bar{R}$-value denoted $\bar{R}_{0}$, and will become negative at $\bar{R}<\bar{R}_{0}$. As was discussed in Ref. \cite{gorbonos}, the surface $\bar{R}=\bar{R}_{0}$ functions as an event horizon, and the vacuum-like solution then corresponds to a BH. [If $h(\bar{R})$ has an additional root at $\bar{R}<\bar{R}_{0}$, it will correspond to an inner horizon.]

\subsection{The Vaidya-like solution}

\label{vaidya} 

Assume now that only one flux vanishes. For concreteness we shall
consider here the \textit{ingoing} Vaidya-like solution, in which 
$\bar{T}_{vv}$ is nonvanishing. Then, as was shown in Ref. \cite{gorbonos}, 
each solution is characterized by a mass-function $\bar{M}(v)$.
The unknown $\bar{R}(u,v)$ is determined by the ODE 
\begin{equation}
\bar{R}_{,v}=h_{0}(\bar{R})-\bar{M}(v).
\label{eq:Rv}
\end{equation}
In turn the other unknown $\bar{S}(u,v)$ is given by 
\begin{equation}
e^{\bar{S}}=-\bar{R}_{,u}.
\label{eq:Sv}
\end{equation}
The nonvanishing influx satisfies $\bar{T}_{vv}=\bar{M}_{,v}$. 
Note that the solution expressed in this from assumes a specific $v$-gauge (to which we may refer as {\it Eddington-like}), but the gauge freedom in $u$ is preserved.  

We shall now derive an expression for $\bar{S}_{,v}$ which was not
included in Ref. \cite{gorbonos}. 
Differentiating $\bar{S}$ from Eq. (\ref{eq:Sv}) yields $\bar{S}_{,v}=\bar{R}_{,uv}/\bar{R}_{,u}$.
Substituting now Eq. (\ref{evf}) for $\bar{R}_{,uv}$ and Eq. (\ref{eq:Sv}) again 
for $\bar{R}_{,u}$, we obtain the simple relation \begin{equation}
\bar{S}_{,v}=-F(\bar{R}).\label{eq:SvNew}\end{equation}
This relation may be useful for constructing approximate
solutions for evaporating $1+1$-dimensional black holes.

The \textit{outgoing} Vaidya-like solution---the analogous situation
in which the nonvanishing flux is $\bar{T}_{uu}$ rather than $\bar{T}_{vv}$---proceeds in a similar manner, though with certain sign changes. One obtains
in this case the expressions $\bar{R}_{,u}=-[h_{0}(\bar{R})-\bar{M}(u)]$,
$e^{\bar{S}}=\bar{R}_{,v}$, $\bar{T}_{uu}=-\bar{M}_{,u}$, and $\bar{S}_{,u}=F(\bar{R})$. 
\footnote{The above relations for both the ingoing and outgoing Vaidya-like solutions hold outside the BH. In the internal region certain sign changes apply, like in the vacuum-like case considered above.}

\section{The CGHS model}

The CGHS model \cite{CGHS} consists of gravity in 1+1-dimensions 
coupled to a dilaton field $\phi$ and to a large number $N$ of identical massless
scalar fields $f_{i}$. CGHS provided an action which includes
an effective semiclassical term derived from the trace anomaly. With
appropriate initial conditions, this model describes the formation
of a BH by a collapsing shell, as well as its subsequent evaporation.
The field equations are most conveniently expressed in double-null
coordinates $u,v$, in which the metric takes the form 
$ds^{2}=-e^{2\rho}dudv$.
The system thus consists of two 
unknowns, $\phi(u,v)$ and $\rho(u,v)$.
The field equations for $\phi$ and $\rho$ are derived
in Ref. \cite{CGHS}. We find it convenient to transform to new variables
$R\equiv e^{-2\phi}$ and $S\equiv2(\rho-\phi)$ (as in Ref. \cite{DO}). The evolution equations
then take the form \begin{equation}
R,_{uv}=-e^{S}-K\rho,_{uv}\ ,\ \ \ S,_{uv}=K\rho,_{uv}/R,\label{evolution_eq_mixed}\end{equation}
where $K\equiv N/12$. 
\footnote{We have set here the cosmological constant
$\lambda=1$ by a trivial shift in $\rho$, as explained in
Ref. \cite{DO}. Also we set the null solution $f_{i}=0$ for the
scalar fields, because we are interested here in the domain of evaporation
where no incoming matter waves are present. The system also contains
two constraint equations, but they are not needed here because any
solution of the evolution equations, with appropriate initial conditions,
automatically satisfies the constraint equations.} 
The Energy-momentum fluxes
$T_{,uu}$ and $T_{,vv}$ are given by 
\begin{equation}
T_{,uu}=R_{,u}S_{,u}-R_{,uu}\:,\;\; T_{vv}\equiv R_{,v}S_{,v}-R_{,vv}. 
\label{Tww}
\end{equation}

The system (\ref{evolution_eq_mixed}) may look quite simple
at first glance, but one should recall that $\rho=(S-\ln\tilde{R})/2$
must be substituted in order to close the system, which messes the right-hand side. Bringing the equations to their standard form, in which $R,_{uv}$ and $S,_{uv}$ are explicitly
given in terms of lower-order derivatives, we end up with the system
\begin{equation}
R,_{uv}=-e^{S}\frac{\left(2R-K\right)}{2\left(R-K\right)}-R,_{u}R,_{v}\frac{K}{2R\left(R-K\right)},
\label{StandardR}
\end{equation}
\begin{equation}
S,_{uv}=e^{S}\frac{K}{2R\left(R-K\right)}+R,_{u}R,_{v}\frac{K}{2R^{2}\left(R-K\right)}.\label{StandardS}
\end{equation}

The classical model is obtained by setting $K=0$, after which the evolution equations reduce to $R,_{uv}=-e^{S},S,_{uv}=0$, which are easily solved. In an Eddington-like gauge the general solution takes the form 
\begin{equation}
R=e^{(v-u)}+f_u(u)+f_v(v) \, \, , \, \, \, S=v-u, 
\label{generalSolution}
\end{equation}
where $f_u(u),f_v(v)$ are arbitrary functions. 
 
 Note that the classical model is a flux-conserving system, with $\bar{R}=R,\bar{S}=S$, and $h_0=\bar{R}$. As such it admits the trivial (vacuum-like) static BH solution 
 $R=e^{(v-u)}+\bar m, S=v-u$, as well as Vaidya-like solutions. 
This straightforward flux-conserving form breaks down once the semiclassical terms are included.

\section{Approximate flux-conserving formulation of CGHS dynamics}

\subsection{Field redefinition}

The presence of terms quadratic in first-order derivatives makes the
field equations (\ref{StandardR},\ref{StandardS}) harder to analyze. We wish to transform the field variables so as to get rid of these terms. A straightforward
algebra leads to the following transformed variables: \cite{bilal}
\begin{equation}
W(R)\equiv\sqrt{R(R-K)}-K\ln(\sqrt{R}+\sqrt{R-K})+K\left(\frac{1}{2}+\ln2\right)\label{eq:Wdef}
\end{equation}
and 
 \begin{equation}
Z\equiv S+\Delta Z(R), \label{eq:Zdef}
\end{equation}
 where 
 \begin{equation}
\Delta Z(R)\equiv\frac{2}{K}(R-W)-\ln R. \label{eq:dZdef}
\end{equation} 
 We shall only be concerned here about the domain $R>K$ 
 (this includes the entire domain-of-dependence portion of the spacetime, both outside and inside the BH, because the singularity is located at $R=K$ \cite{DO}). 
  Since $dW/dR=\sqrt{1-K/R}$, the function $W(R)$ is monotonically increasing throughout this domain. Therefore, the inverse function $R(W)$ is mathematically well defined, though it cannot be expressed explicitly. 

With these new variables the evolution equations take the
schematically-simpler potential form:
\begin{equation}
W_{,uv}=e^{Z}\,V_{W}(W)\;,\quad Z_{,uv}=e^{Z}\,V_{Z}(W), 
\label{potentialForm}
\end{equation}
with certain {}``potentials'' $V_{W}(W)$ and $V_{Z}(W)$. But the simplification 
does not come without a cost: These potentials are explicitly obtained as functions of $R$ rather than $W$. One
finds 
\begin{equation}
V_{W}=-\frac{R-K/2}{\sqrt{R(R-K)}}\, e^{-\Delta Z(R)}\label{eq:VWdef}
\end{equation}
 and
 \begin{equation}
V_{Z}=\frac{2}{K}\left[\frac{R-K/2}{\sqrt{R(R-K)}}-1\right]\, e^{-\Delta Z(R)}.\label{eq:VZdef}
\end{equation}
Nevertheless, the potential form (\ref{potentialForm}) is useful for various analytical considerations, despite its implicit character. Furthermore, we shall
primarily be concerned here about the leading-order semiclassical effects, 
and as it turns out, these leading-order  effects are fully encoded in the large-$R$ (or, equivalently, large-$W$) asymptotic form of the field equations.  
At the desired order the functions $V_{W}(W)$ and $V_{Z}(W)$ [as well as $R(W)$] all take simple explicit forms, as we shortly demonstrate.

\subsection{Large-$R$ asymptotic behavior}

The semiclassical effects, responsible for the BH evaporation, are
encoded in the $K$ terms in the field equations (\ref{StandardR},\ref{StandardS}) or (\ref{eq:Wdef}-\ref{eq:VZdef}). 
(Notice that $W$ and $Z$ respectively approach  $R$ and $S$ as $K\to 0$, and the field equations then reduce to their classical form,
namely $V_W=-1,V_Z=0$.) To depict the
leading semiclassical effect we may formally decompose all relevant functions in powers of $K$. Doing so, one observes that this effectively becomes an expansion in $K/R$ (see below). Although $K$ itself may be a large number, as long as the BH is macroscopic (namely, its mass is $\gg K$), $K/R\ll 1$ is satisfied  throughout the BH exterior (and also in the interior, except in the neighborhood of the $R=K$ singularity---a region which will not concern us here). Thus, to depict the
leading-order semiclassical effect we decompose all relevant quantities to
first order in $K/R$ (or, equivalently, $K/W$). For the 
transformation of variable we obtain 
\begin{equation}
W=R\left[1-\frac{K}{2R}\ln R+O\left(\frac{K}{R}\right)^{2}\right] 
\end{equation}
and 
\begin{equation}
\Delta Z=-\frac{K}{4R}+O\left(\frac{K}{R}\right)^{2}. 
\end{equation}
The inverse function $R(W)$ is given at this order by 
\begin{equation}
R=W\left[1+\frac{K}{2W}\ln W+O\left(\frac{K}{W}\right)^{2}.\right] 
\end{equation}
The potentials $V_{W}(W),V_{Z}(W)$ can now easily be expanded to
first order in $K/W$, 
\begin{equation}
V_{W}=-1-\frac{K}{4W}+O\left(\frac{K}{W}\right)^{2}\, \,  ,
\, \, \, \, \, 
V_{Z}=\frac{1}{W}\left[\frac{K}{4W}+O\left(\frac{K}{W}\right)^{2}\right].\label{eq:VZaprx}
\end{equation}
Recall that this leading-order semiclassical approximation is applicable as long as $K/R\approx K/W\ll 1$, which for a macroscopic BH holds everywhere in the exterior (and also throughout most of the BH interior).

\subsection{The large-$R$ flux-conserving system}

We now proceed with the above large-$R$ approximation, and analyze the PDE 
system
\begin{equation}
W_{,uv}=e^{Z}V_{W}(W)\;,\quad Z_{,uv}=e^{Z}V_{Z}(W)\label{eq:system}\end{equation}
 with 
 \begin{equation}
V_{W}=-1-\frac{K}{4W}\; ,\quad V_{Z}=\frac{K}{4W^{2}}\; ,
\label{eq:VWZ}
\end{equation}
omitting all higher-order terms $\propto (K/W)^2$. 

Since $V_{Z}=dV_{W}/dW$, this is a flux-conserving system: 
It is described by Eq. (\ref{evf}), with $\bar{R}=W$, $\bar{S}=Z$,
and $F=V_{W}$. Its generating function is 
\begin{equation}
h_{0}(W)=W+\frac{K}{4}\ln W. 
\label{eq:h(W)}
\end{equation}

In particular the system (\ref{eq:system},\ref{eq:VWZ}) admits Vaidya-like
solutions. As it turns out \cite{OriPrep}, at the leading order an evaporating BH may be approximated by an ingoing Vaidya-like solution 
with a linearly-decreasing mass function. 
\footnote{However, a weak outflux term $\propto (K/W)$ has to be added in order to match the initial conditions, as discussed below.} 
Assuming that the BH was created by the collapse of
a shell of mass $M_{0}$, which propagated
along the ingoing null orbit $v=v_{0}$, the mass function takes the form $\bar{M}(v)=M_{0}-(K/4)(v-v_{0})$.
By shifting the origin of $v$ such that $v_{0}=-(4/K)M_{0}$, we
obtain \begin{equation}
\bar{M}(v)=-\frac{K}{4}v\equiv m_{v}(v).\label{eq:mv}\end{equation}
Note that $dm_{v}/dv=-K/4=-N/48$, which reflects the standard evaporation
rate of a two-dimensional BH \cite{CGHS}. 

The determination of the evaporating-BH spacetime thus reduces within this approximation to analyzing the ODE (\ref{eq:Rv}) which now reads 
\begin{equation}
W_{,v}=h_{0}(W)-\bar{M}(v)=(W+\frac{K}{4}\ln W)+\frac{K}{4}v.\label{eq:Wv}
\end{equation}
 The other unknown $Z$ is then obtained from Eq. (\ref{eq:SvNew}),
which reads 
\begin{equation}
Z_{,v}=1+\frac{K}{4W}. 
\label{eq:Zv} 
\end{equation}
[Alternatively $Z$ may be obtained from Eq. (\ref{eq:Sv}).]
The dependence upon $u$ enters through the initial conditions
at $v=v_{0}$ (the collapsing shell). The appropriate initial conditions
at the shell will be discussed in Ref. \cite{OriPrep}. 

Let us denote the solution to the two ODEs (\ref{eq:Wv},\ref{eq:Zv})
by $W_{c}(u,v)$ and $Z_{c}(u,v)$. Although it constitutes an exact
solution to the system (\ref{eq:system},\ref{eq:VWZ}), as it stands it does not
yield an adequately precise description of the CGHS BH spacetime. 
Indeed an approximation was involved in the first place in deriving the flux-conserving system (\ref{eq:system},\ref{eq:VWZ}) from the original CGHS system.
This approximation is fine, because the terms omitted were of order
$(K/W)^{2}$, whereas our attempt here is to construct an approximate solution
at order $K/W$. As it turns out \cite{OriPrep}, however, the solution $(W_{c},Z_{c})$ fails to satisfy the proper initial conditions at the shell ($v=v_{0}$), and the discrepancy is at order $K/W$, which cannot be neglected here. Stated in other words,
the flux-conserving solution which properly describes the evaporating
BH is not a pure ingoing Vaidya-like solution, but one which
also contains a weak outflux---namely, a small correction $\delta W(u) \propto K$ to $W=W_c$. 
Nevertheless, owing to the smallness of $\delta W$ 
(and to the trivial form of the field equations at the limit $K\to 0$), 
one can simply superpose the correction $\delta W(u)$ on top of the Vaidya-like solution $W_{c}$ (with $Z_{c}$ unchanged) \cite{OriPrep}. 
This procedure yields a sufficiently accurate solution which satisfies the appropriate initial conditions. 
\footnote{In fact it is the superposed outgoing component $\delta W(u)$ which gives rise to the Hawking outflux at future null infinity.}

In a forthcoming paper \cite{OriPrep} we shall implement this formalism, and construct an approximate solution $R(u,v),S(u,v)$ to the CGHS field equations which satisfies the correct initial conditions (at the shell as well as at past null infinity), and is accurate at first order in $K/W$.

\end{document}